\documentclass[10pt]{article}
\pdfoutput=1
\usepackage{graphicx,amssymb,amsfonts,amsmath,amssymb,amscd,amstext, mathrsfs,color}
\makeatletter \@addtoreset{equation}{section} \makeatother

\newcommand{\be}{\begin{eqnarray}}
\newcommand{\ee}{\end{eqnarray}}
\newcommand{\ba}{\begin{array}}
\newcommand{\ea}{\end{array}}
\newcommand{\bal}{\begin{align*}}
\newcommand{\eal}{\end{align*}}

\newcommand{\nn}{\nonumber}

\renewcommand{\(}{\Big(}
\renewcommand{\)}{\Big)}

\def \<{\langle}
\def \>{\rangle}
\definecolor{ggg}{rgb}{0,.6,0}
\begin{document}
~
\vspace{0.5cm}
\begin{center} {\Large \bf Third-Order Perturbative OTOC of the Harmonic Oscillator with Quartic Interaction and Quantum Chaos}
\\
                                            
\vspace{1cm}

Wung-Hong Huang*\\
\vspace{0.5cm}
Department of Physics, National Cheng Kung University,\\
No.1, University Road, Tainan 701, Taiwan\\                    
\end{center}
\begin{center} {\large \bf  Abstract} \end{center}
We calculate the third-order out-of-time-order correlator (OTOC) of a simple harmonic oscillator with an additional quartic interaction using the second quantization method.  We obtain analytic relations for the spectrum, Fock space states, and matrix elements of the coordinate, which are then used to numerically evaluate the OTOC.  We observe that after the scrambling, the OTOC becomes a fluctuation around a saturation point at later times, which is associated with quantum chaotic behavior in systems that exhibit chaos. We analyze the early-time properties of the OTOC and find that in systems with sufficiently strong quartic interactions, an exponential growth curve fitting over a long time window clearly emerges in the third-order perturbation 
\\
\\
\scalebox{0.6}{\hspace{1cm}\includegraphics{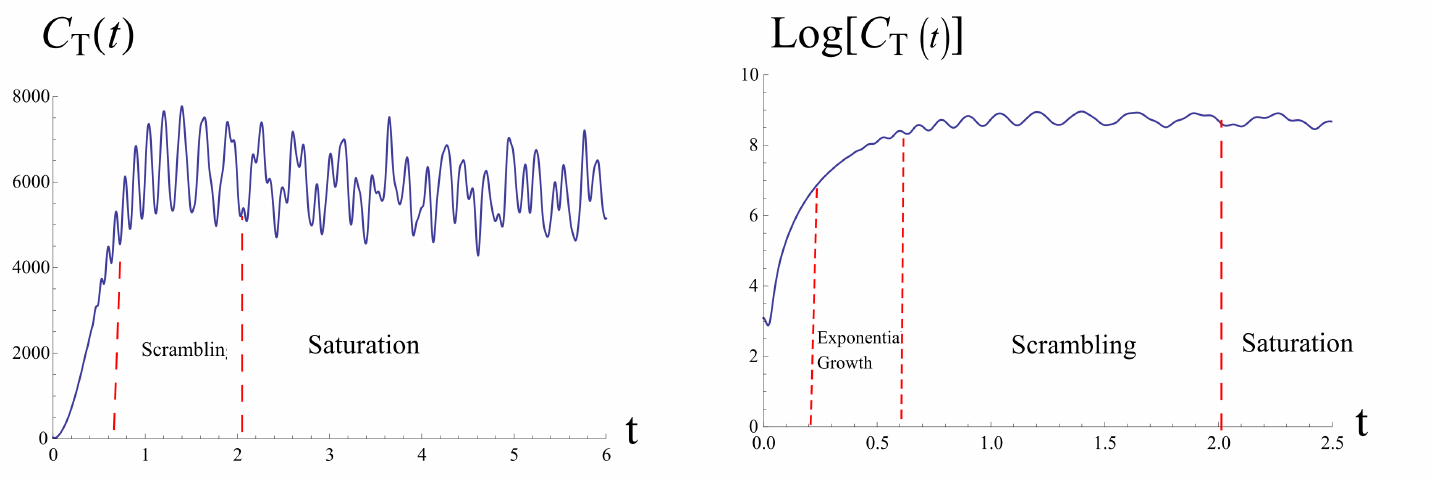}}
\\
\scalebox{0.6}{\hspace{1cm}\includegraphics{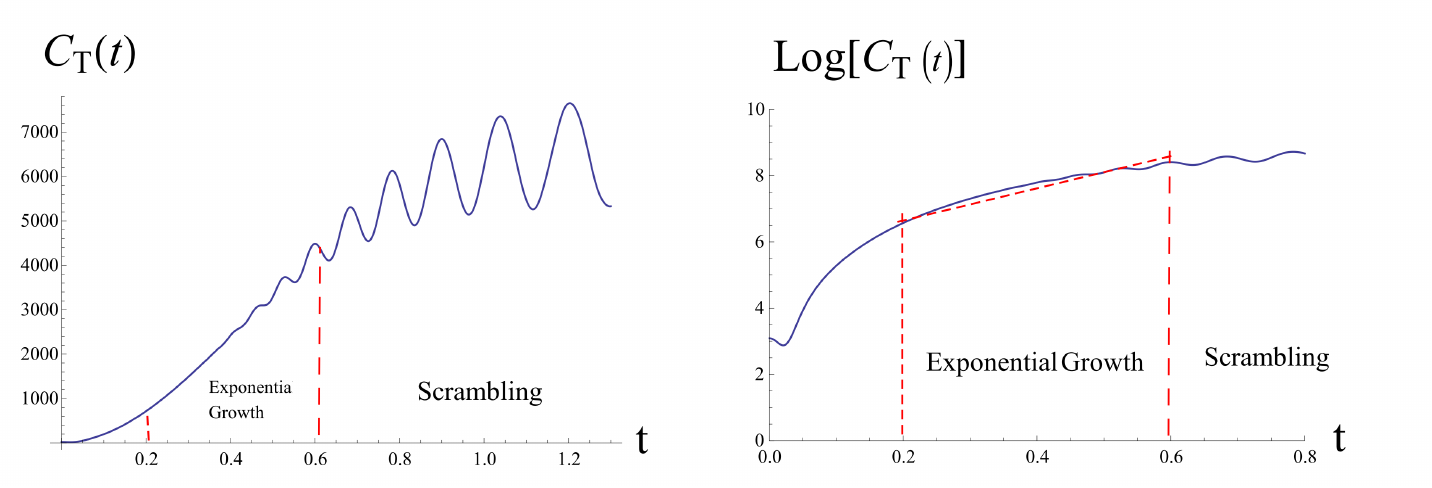}}
\begin{flushleft}
* Retired Professor of NCKU, Taiwan. \\
* E-mail: whhwung@mail.ncku.edu.tw
\end{flushleft}
\newpage
            \tableofcontents
\section{Introduction}
The  out-of-time-order correlator (OTOC) is known to be an important quantity to indicate   quantum chaos after the  associated exponential growth property was discussed by  Larkin and Ovchinnikov   many years ago \cite{Larkin}  . Since  it was revived by  Kitaev \cite{Kitaev15a,Kitaev15b,Sachdev}  in recent years,  the OTOC has attracted significant attention in the physics community across various fields, including condensed matter physics and  high-energy physics.  Particularly, after  the discovery that the Lyapunov exponent  saturates a bound \cite {Maldacena15,Maldacena16, Kitaev15c} many researchers have been drawn to study problems related to conformal field theory and AdS/CFT duality. \cite{Shenker13a, Shenker14, Roberts14,  Shenker13b,Susskind,Liam,Verlinde,Kristan}.  

The function of out-of-order correlator (OTOC)  is defined by
\be
 C_T(t)=-\<[W(t),V(0)]^2\>_T   \sim  e^{2\lambda t}
\ee
In the case of  W(t) = x(t)  and V = p the quantity  $C_T(t)  =\hbar^2({\partial x(t)\over \partial x(0)})^2 $  can be obtained by using the  classical-quantum correspondence.  We define the Lyapunov exponent  $\lambda$ by $|{\partial x(t)\over \partial x(0)}|\sim  e^{\lambda t}$, which  measures    sensitivity to initial conditions.  Consequently,  the quantum OTOC  grows as $\sim  e^{2\lambda t}$. 
 Thus, we can extract the quantum Lyapunov exponent $\lambda$  from OTOC.   Maldacena, Shenker, and Stanford \cite{Maldacena15} found from gravity side  that  the Lyapunov exponent is bounded by temperature T : $ \lambda \le  {2\pi  T}$. In this context, several studies have explored the problem under external fields or in higher gravity theories, as discussed in  \cite{Andrade,Sircar, Kundu,Ross,Huang16,Huang17,Huang18}.

The quantum mechanical method of calculating OTOC  with general Hamiltonian  was set up by Hashimoto recently in  \cite{Hashimoto17,Hashimoto20a,Hashimoto20b}.  For  simple  harmonic oscillator (SHO) the OTOC can be calculated exactly and  is  a purely oscillatory function.  

Along the method, many complicated examples were studied, such as the two-dimensional stadium billiard \cite{Hashimoto17,Rozenbaum2019}, the Dicke model \cite{Chavez-Carlos2018}, and bipartite systems \cite{Prakash2020}. These models can exhibit classical chaos, where numerical calculations show that OTOCs grow exponentially at early times, followed by saturation at late times. The method has also been applied to study several systems, including many-body physics, as seen in \cite{Das, Romatschke, Morita, Swingle}.
\\

In this framework,  the properties of the OTOC   are primarily determined using numerical methods in the wave function approach. In a  previous  note  \cite{Huang2306} we began to   study the OTOC of  a simple harmonic  oscillator with extra anharmonic (quartic) interaction by an analytical perturbation method in second  quantization approach\footnote{The reference \cite{Romatschke} examined the OTOC of oscillators with pure quartic interaction in wavefunction approach, where the system has an exact solution for both  the wave function and the spectrum.}.  The perturbation method has the advantage of allowing us to find the properties of any quantum level "n", whereas in the wave function approach, one can obtain the properties of the quantum level "n" only after performing numerical evaluations step by step for each level .

According to our method, to the first order pertubation of OTOC \cite{Huang2306}, however, we does not see the exponential growth in the initial time nor a fluctuation around a saturation point at later times.  In a  subsequent  paper  \cite{Huang2311} we extended the method to the second-order perturbation and found that the OTOC exhibits a fluctuation around a saturation point at later times. However, the obtained analytic formula shows  that   at early times the OTOC will rapidly raise in the quadratic power law, not the exponential growth, which   is essential for the emergence of chaotic dynamic. We present the properties  of $\log [C_T(t)]$ for the cases of first-order and second-order perturbations in Figure 1, which is to be compared with Figure 2 for the case of third-order perturbation studied in this paper. 
\\
\\
\scalebox{0.6}{\hspace{2cm}\includegraphics{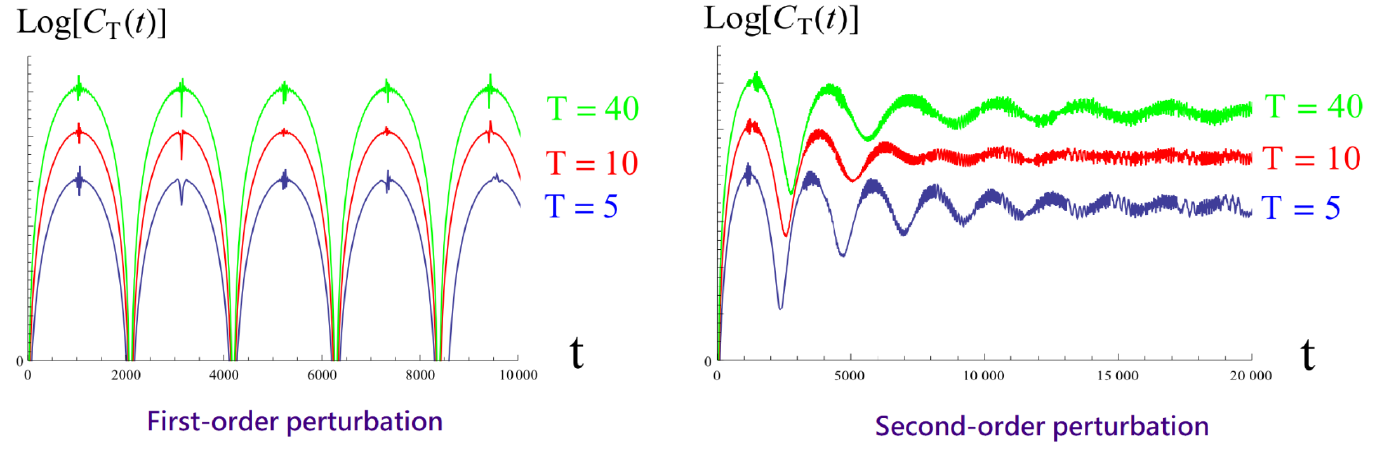}}
\\
{Figure 1:  First and second orders   OTOC   $C_T(t)$ as a function of time.}
\\

In this paper, we will present evidence that the exponential growth of the OTOC can be shown in third-order perturbation.   In Section 2, we briefly  review  Hashimoto's method  for  calculating quantum mechanic  OTOC  in the  simple harmonic oscillator and then  use the second quantization method to obtain the same result   quickly.  In Section 3, we  use the   second quantization method to calculate the OTOC in the systems  of  harmonic  oscillator with extra anharmonic (quartic) interation.  We obtain analytic formulas  for the  spectrum, Fock space states, and matrix elements of the coordinate to the third-order   of anharmonic interaction.  Using these relations, in Section 4  we numerically calculate the thermal    OTOC and analyze  its properties. We show that  in  systems with sufficiently strong quartic interactions, an exponential growth curve fitting over a long time window clearly emerges  in the third-order perturbation. The system then scrambles and exhibits fluctuations around a saturation point at later times. The last section is devoted to brief discussions and mentions some directions for future studies.
\section{OTOC in Quantum Theory }
\subsection{Quantum Mechanic Approach to OTOC}
We first briefly review the  Hashimoto's method of calculating  OTOC in quantum mechanic  model \cite{
 Hashimoto17}\footnote{Refer to \cite{Huang2311} for a more detailed review.}.     For a  time-independent Hamiltonian: $H = H(x_1,....x_n,p_1,....p_n)$ the function of  OTOC is  
\be
C_T(t)=-\<[x(t),p(0)]^2 \>_T  
\ee
Using energy eigenstates $|n\>$,  defined by $H|n\>=E_n|n\>$ then
\be
  C_T(t)& =&{1\over Z}\sum_ne^{-\beta E_n}\,c_n(t)  ,~~~c_n(t)\equiv -\<n|  [x(t),p(0)]^2         |n\>    \label{TC}\\
  c_n(t)   & =&    \sum_m(ib_{nm})(ib_{nm})^*  ~~~b_{nm}= -i\<n|  [x(t),p(0)]|m\>,~~b_{nm}^*=b_{mn}  ~\label{cn}
\ee
Substituting a relation $ x(t) = e^{ iHt/\hbar  }\,x\,e^{- i  Ht/\hbar} $  we    obtain
\be
 b_{nm}&\equiv& -i\<n|  x(t), p(0)|m\> +i\<n| p(0) x(t),|m\>\nn\\
&=& -i\sum_k\(e^{i  E_{nk}t/\hbar}x_{nk}p_{km}-e^{i E_{km}t/\hbar}p_{nk}x_{km}\) \\
E_{nm}&=& E_n-E_m,~~~ x_{nm}=\<n| x|m\> ,~~p_{nm}=\<n|p |m\>~~~~\label{Enm}
\ee
For  the quantum mechanical Hamiltonian\footnote{{Note that Hashimoto \cite{Hashimoto17} used $H=\sum_i{p_i^2}+U(x_1,....x_N)$ which is that in our notation for M=1/2.  Therefore the formula $b_{mn}$ in eq.(\ref{bnm})  becomes Hashimoto's   formula if M=1/2 and $\hbar=1$.}}
\be
&&H=\sum_i{p_i^2\over 2M}+U(x_1,....x_N)~~\to~~[H,x_i] = - i\hbar { p_i\over M} 
\ee
where  $M$ is the particle mass. The relations $p_{km}=\<k|p |m\>={iM\over \hbar } E_{km}  $ leads to  
\be
 b_{nm} = { M\over \hbar }\sum_k\,x_{nk}x_{km}\(e^{i  E_{nk}t/\hbar} E_{km} -e^{i  E_{km}t/\hbar}E_{nk}\)  ~~~~~~~\label{bnm}
\ee
and  we can compute OTOC through (\ref{bnm}) once we know $x_{nm}$ and $E_{nm}$ defined in (\ref{Enm}).
\subsection{OTOC of SHO}
Using the above formula, let's consider the example of calculating the OTOC of SHO. The Hamiltonian  $H$, spectrum $E_n$, and state wavefunction $\Psi_n(x)$ have exact forms in any textbook of quantum mechanics. Therefore  $ E_n=\hbar\omega\left(n+\frac{1}{2}\right)$, $E_{nm}=\hbar\omega(n-m)$, and
\be
 x_{nm}&=&\<\Psi_n(x)| \,x\, |\Psi_m(x)\>= \sqrt{\hbar\over  2M\omega}\, \(\sqrt{n }  \ \delta_{n,m+1}+\sqrt{n+1 }  \ \delta_{n,m-1}\)   ~~~~~~\label{xnm}
  \ee
where $ n,m=0,1,2,\cdots $.  Substituting above expressions into  (\ref{Enm}) and  (\ref{bnm}) we obtain 
\be 
 b_{nm}(t) & {=}&  \hbar \cos (\omega t)\ \delta_{nm}   \label{first}      \\
\Rightarrow~
c_n(t)&=&   \hbar^2  \cos^2 (  \omega t),~~~C_T(t)=  \hbar^2 \cos^2(  \omega t) ~~~~~~~~~~~\label{HR}
\ee
It is seen that both of $c_T(t)$ and $C_T(t)$ are periodic functions and   do not depend on energy level $n$ nor temperature $T$. This is a special property   for the harmonic oscillator.

We can use  the second quantization  to obtain above result quickly.  In the second quantization the state  is denoted as $|n\>$. The   creation and destroy operators have the propery : $a^\dag|n\>= \sqrt{n+1} |n+1\>,~~~  a|n\>= \sqrt{n} |n-1\>$.  Applying above  relations and following definitions
\be
x&=&\sqrt{\hbar\over2M\omega}(a^\dag+a ),~~~p= i\sqrt{M\omega\hbar\over2}(a^\dag-a) \\
\Rightarrow x_{nm}&\equiv&\<n|x|m\>=\sqrt{\hbar\over2M\omega}\( \sqrt{m}\ \delta_{n,m-1}+   \sqrt{m+1}\ \delta_{n, m+1}\)   \label{SQxnm}
\ee
which exactly reproduces (\ref {xnm}) and, therefore the values of  $b_{nm}$, $c_n(t)$ and $C_T(t)$ in (\ref{HR}).   
%
\section{Third-order Perturbative OTOC of Anharmonic Oscillator : Analytic Relations}
In this section  we will calculate the third-order perturbative OTOC of  standard simple harmonic  oscillator while with extra quartic interaction   in the second quantization.  Note that first-order and second-order calculations had been completed in our previous notes \cite{Huang2306} and \cite{Huang2311} respectively. While the third-order calculations in this paper are relatively complex the results could show the exponential growth  properties.
\subsection{Perturbative Energy and State : Anharmonic Model and Formulas}
We consider the standard simple harmonic  oscillator while with extra  quartic  interaction 
\be
 H&=&({p^2\over 2M}+{M\omega^2\over 2}x^2)+g{x^4}\nn\\
&=&  \hbar\omega \( a^\dag a+ \frac{1}{2}\) +{g } \ \({\hbar\over2M\omega}\)^2(a^\dag+a )^4=H^{(0)} +gV
\ee
which has  a well-known unperturbed solution
\be
H^{(0)} |n^{(0)} \>=E^{(0)}_n |n^{(0)} \>=\hbar\omega\(n^{(0)} +{1\over2}\) |n^{(0)}  \> 
\ee

To third-order perturbation the  energy and the state formulas in quantum mechanics  are 
\be
E_n &= & E_n^{(0)} + g E_n^{(1)} + g^2 E_n^{(2)} + g^3E_n^{(3)}+{\cal O}(g^4) \\
 |n\>&=& |n^{(0)} \>+g  |n^{(1)} \>+g^2 |n^{(2)} \>+g^3 |n^{(3)} \>+{\cal O}(g^4)            \label{n}
 \ee
where
\be
 E_n^{(1)} &=&  V_{nn} =\<n^{(0)} | V   |n^{(0)}  \>,~~~ ~~~~           \\ 
  E_n^{(2)} &=&   { V_{nk }^2   \over E_{nk }} =    \sum_{k\ne n}{  \<n^{(0)} |V|k^{(0)} \>  ^2   \over E^{(0)}_n -E^{(0)}_k}                         \\         
 E_n^{(3)} &=&  -V_{nn }{ V_{nk }^2   \over E_{nk }^2}  +      { V_{nk_1 } V_{k_1k_2 } V_{k_2n }   \over E_{nk_1 }E_{nk_2 }}   \\
&=&  -      \<n^{(0)} | V   |n^{(0)}  \>\sum_{k\ne n}  { V_{nk }^2   \over E^2_{nk }}+\sum_{k_1\ne n}\sum_{k_2\ne n}{  \<n^{(0)} |V|{k_1}^{(0)} \> \<{k_1}^{(0)} |V|{k_2}{(0)} \> \<{k_2}^{(0)} |V|n^{(0)} \>      \over (E^{(0)}_n -E^{(0)}_{k_1})(E^{(0)}_n -E^{(0)}_{k_2}) } ~~
\ee
and, using above notion we have relations
\be
 |n^{(1)} \> &=&{V_{kn}\over E_{nk}} |k^{(0)}\>   \\
 |n^{(2)} \>&=&  \left( {V_{k_1k_2} V_{k_2n}\over E_{nk_1} E_{nk_2}}-  {V_{nn} V_{k_1n}\over E_{nk_1}^2} \right)  |k_1^{(0)}\>       -{1\over2}{V_{nk} V_{kn}\over E_{nk}^2}     |n^{(0)}\>            \\ 
|n^{(3)} \>&=&  \left[ -   { V_{k_1k_2 } V_{k_2 k_3 }  V_{  k_3n }  \over E_{k_1n }E_{nk_2 }E_{nk_3 }}     + { V_{nn } V_{k_1 k_2 }  V_{  k_2n }  \over E_{k_1n }E_{nk_2 }}\( {1\over  E_{nk_1 }}    +          {1\over  E_{nk_2 }}\) - {V_{nn}^2 V_{kn}\over E_{kn}^3}+ { V^2_{nk_2 } V_{k_1n } \over E_{k_1n }E_{nk_2 }}\( {1\over  E_{nk_1 }}+ {1\over 2 E_{nk_2 }} \) \right] | k^{(0)  } \> \nn\\
&&+\left[ -   { V_{k_1n} V_{nk_2 } V_{k_2k_1 }+V_{nk_1} V_{k_1k_2 } V_{k_2 n }  \over  E^2_{nk_2 }E_{nk_1 }}+{V_{nk}^2 V_{nn}\over E_{kn}^3}\right]| n^{(0)  } \>
\ee
In the rest of this section we will use   above formulas to calculate: 

1.  Perturbative Energy   $  E$. 

2. Perturbative state $ {|n\>}$. 

3. Perturbative matrix elements $ {x_{mn}}$. 
\\
 Using these analytic results we will  calculate  the OTOC and analyze its property in the next section.
\subsection{Perturbative Energy $  E_n$  : Model Calculations}
 We   use the unit : $\hbar=\omega=M=1$ and keep the coupling strength $g$ as a only free parameter. A crucial quantity we need, after calculation, is
\be
V_{kn}&=&\<k^{(0)})|V|n^{(0)}\>\nn\\
&=& \frac{  \sqrt{n-3} \sqrt{n-2} \sqrt{n-1} \sqrt{n} }{4 }\ \delta _{k,n-4}+\frac{  \sqrt{n-1} \sqrt{n}\ (2 n-1)   }{2  }\ \delta _{k,n-2} +\frac{ 3 (2 n (n+1)+1)   }{4 }\ \delta _{k,n }\nn\\
&&+\frac{  \sqrt{n+1} \sqrt{n+2}\ (2 n+3)   }{2 }\ \delta _{k,n+2}  +\frac{  \sqrt{n+1} \sqrt{n+2} \sqrt{n+3} \sqrt{n+4}   }{4  }  \ \delta _{k,n+4}
\ee
Using above result the third-order perturbative   energy $ {E_n}$  becomes
\be
E_n &= & E_n^{(0)} + g E_n^{(1)} + g^2 E_n^{(2)} + g^3E_n^{(3)}+{\cal O}(g^4) \\
 {E^{(0)}_n} &=& \(n+{1\over2}\)   \\
{E^{(1)}_n} &=&{3 \over 4} (1 + 2 n (1 + n))      \\
{E^{(2)}_n} &=&- {1 \over 8} ((1 + 2 n) (21 + 17 n (1 + n)))    \\
 {E^{(3)}_n} &=&   \frac{ 1}{512} \left(11748 + n (1 + n) (37202 + n (1 + n) (14987 + 390 n (1 + n)))\right)                       ~~\label{NE} \\
   {E_{nm}} & =& {E_n} -  {E_m}
\ee
This relation shows an interesting  property of   “enhancement" which we mention  it in below.
\subsubsection{Enhancement Property}  \label{Enhancement}
  In the case of  $n> 1$ above relation leads to 
\be
  {E_n}  &\approx & n( \alpha_0  + \alpha_1 g   n+ \alpha_2 g^2  n^2+ \alpha_3 g^3  n^5) ~~~~~\label{crucial}
\ee
and we see that, for example,  for small value of  $g=0.01$  the value of $gn$ is larger then $1$  if the mode of  state $n>100$. This leads to a general property of  “enhancement"  that no matter how  small the value of coupling strength $g$ is, some perturbative quantities can be significantly large in higher energy levels. To ensure the reliability of perturbation theory, it is necessary to study systems with low-energy levels. This constraint leads to the consideration of systems at low temperatures, as higher energy levels will be suppressed by the Boltzmann factor. The numerical investigations in the next section are at T=60 for this reason.
 
\subsection{Perturbative  State $ {|n\>}$   : Model Calculations}
To proceed we   calculate the perturbative  states of $  {|n\>} $.  The results, with a notation $|n^{(j)}\>=|n \>^{(j)}$, are
\be
{|n\>}&=& |n^{(0)} \>+g  |n^{(1)} \>+g^2 |n^{(2)} \>+g^3 |n^{(3)} \>+{\cal O}(g^4) \\
\nn\\
|n\>^{(1)}&=&\frac{1}{16} \sqrt{n-3} \sqrt{n-2} \sqrt{n-1} \sqrt{n} \ |n-4\>^{(0)}+\frac{1}{4} \sqrt{n-1} \sqrt{n} (2 n-1)\ |n-2\>^{(0)}\nn\\
&&-\frac{1}{4} \sqrt{n+1} \sqrt{n+2} (2 n+3)  \ |n+2\>^{(0)}-\frac{1}{16} \sqrt{n+1} \sqrt{n+2} \sqrt{n+3} \sqrt{n+4} \ |n+4\>^{(0)}\nn\\   \label{1} \\
|n \>^{(2)}&=&\frac{1}{512}   \sqrt{n-7} \sqrt{n-6} \sqrt{n-5} \sqrt{n-4} \sqrt{n-3} \sqrt{n-2} \sqrt{n-1} \sqrt{n}\ |n-8\>^{(0)}\nn\\
&&  +\frac{1}{192}   \sqrt{n-5} \sqrt{n-4} \sqrt{n-3} \sqrt{n-2} \sqrt{n-1} \sqrt{n} (6 n-11)
 |n-6\>^{(0)}   \nn\\
&&+\frac{1}{16}   \sqrt{n-3} \sqrt{n-2} (n-1)^{3/2} \sqrt{n} (2 n-7)  \ |n-4\>^{(0)}    \nn\\
&&+\frac{1}{64}   \sqrt{n-1} \sqrt{n} (n (n (2 n+129)-107)+66)\ |n-2\>^{(0)}  \nn\\
&&-\frac{1}{256}   (n (n+1) (65 n (n+1)+422)+156)  \ |n \>^{(0)}\nn\\ 
&& +\frac{1}{64}   \sqrt{n+1} \sqrt{n+2} (n (n (123-2 n)+359)+300)  \ |n+2\>^{(0)}    \nn\\
&& +\frac{1}{16}   \sqrt{n+1} (n+2)^{3/2} \sqrt{n+3} \sqrt{n+4} (2 n+9) \ |n+4\>^{(0)}          \nn\\  
&& +\frac{1}{192}   \sqrt{n+1} \sqrt{n+2} \sqrt{n+3} \sqrt{n+4} \sqrt{n+5} \sqrt{n+6} (6 n+17) \ |n+6\>^{(0)}          \nn\\
&& +\frac{1}{512}   \sqrt{n+1} \sqrt{n+2} \sqrt{n+3} \sqrt{n+4} \sqrt{n+5} \sqrt{n+6} \sqrt{n+7} \sqrt{n+8}  \ |n+8\>^{(0)}    \label{2}    
\ee
Function  Form  of $|n\>^{(3)}$ is expressed in appendix A.
\subsection{Perturbative  Matrix Elements $ {x_{mn}}$ : Model Calculations}
Use above relations we could now begin to  calculate the   matrix elements $ {x_{mn}}=\<m|x|n\>$. First, we write 
\be
|n \>^{(j)} &=&  \sum_{k=\{k_j\}} f_j(n,k) | n+k   \>^{(0)},~~~~~j=1,2,3            \label{ni}\\
 \{k_1\} &=&  \pm4,\pm2;~\{k_2\}=\pm8,\pm6,\pm4\pm2,0;~\{k_3\}=\pm12,\pm10,\pm8,\pm6,\pm4\pm2,0 
\ee
where the functions $f_j(n,k)$ could be read from  eq.(\ref{1}) $\sim$ eq.(\ref{3}).  

To third order of $g$ the state ${|n\>}$ and  $x|n\>$ becomes
\be 
{|n\>}&=&  |n \>^{(0)}+g  |n \>^{(1)}+g^2 |n\>^{(2)}+g^3 |n \>^{(3)} \nn \\
&=& |n \>^{(0)}+   \sum_{k=\{k_1\}} g f_1(n,k) | n+k   \>^{(0)}+ \sum_{k=\{k_2\}} g^2 f_2(n,k) | n+k   \>^{(0)}+  \sum_{k=\{k_3\}} g^3 f_3(n,k) | n+k   \>^{(0)}  \nn\\
\ee
\be
x|n\>&=&{1\over \sqrt 2}(a^\dag+a)|n\>={1\over \sqrt 2}(a^\dag+a)  \(|n \>^{(0)}+g  |n \>^{(1)}+g^2 |n\>^{(2)}+g^3 |n \>^{(3)}\)   \nn\\
& =& {1\over \sqrt 2} \(\sqrt{n+1} \ |n+1 \>^{(0)}+\sqrt{n}\  |n-1 \>^{(0)}\) \nn\\
&&+ {g\over \sqrt 2}  \sum_{k=\{k_1\}}    f_1(n,k)   \(\sqrt{n+k+1} \ |n+k+1 \>^{(0)}+\sqrt{n+k}\  |n+k-1 \>^{(0)}\)\nn   \\
&&+ { g^2\over \sqrt 2}  \sum_{k=\{k_2\}}  f_2(n,k)  \(\sqrt{n+k+1} \ |n+k+1 \>^{(0)}+\sqrt{n+k}\  |n+k-1 \>^{(0)}\)\nn   \\
&&+ { g^3\over \sqrt 2}  \sum_{k=\{k_3\}}  f_3(n,k)  \(\sqrt{n+k+1} \ |n+k+1 \>^{(0)}+\sqrt{n+k}\  |n+k-1 \>^{(0)}\)                     \label{3.24}
\ee
 Therefore
\be
\<m|x|n\>= \({}^{(0)}\<m|+g~{}^{(1)}\<m|+ g^2~ ^{(2)}\<m|+ g^3~ ^{(3)}\<m| \)                                 \        x\   \(|n \>^{(0)}+g  |n \>^{(1)}+g^2 |n\>^{(2)}+g^3 |n \>^{(3)}\) ~~ 
\ee
and   the third order matrix elements  $  {\<m|} x {|  n   \>} $ becomes
\be
{\<m|} x {|  n   \>} &=&  \frac{g^3 \sqrt{n+1} \sqrt{n+2} \sqrt{n+3} \sqrt{n+4} \sqrt{n+5} \sqrt{n+6} \sqrt{n+7}}{64 \sqrt{2}} \  \delta_{m,n+7}                \nn\\
&& +  \frac{g^2 \sqrt{n+1} \sqrt{n+2} \sqrt{n+3} \sqrt{n+4} \sqrt{n+5} (73 g (n+3)-4) }{64 \sqrt{2}}                                             \  \delta_{m,n+5}    \nn\\
&& +     \frac{g \sqrt{n+1} \sqrt{n+2} \sqrt{n+3} \left(g^2 \left(3219 n^2+12876 n+14041\right)-312 g (n+2)+32\right)  }{128 \sqrt{2}}   \  \delta_{m,n+3}                                           \nn\\
&& +   \frac{\sqrt{n+1} \left(-3 g^3 \left(842 n^3+2526 n^2+3193 n+1509\right)+g^2 \left(303 n^2+606 n+378\right)-48 g (n+1)+32\right)}{32 \sqrt{2}}   \  \delta_{m,n+1}                                             \nn\\
&& +    \frac{\sqrt{n} \left(-3 g^3 n \left(842 n^2+667\right)+g^2 \left(303 n^2+75\right)-48 g n+32\right)}{32 \sqrt{2}}       \  \delta_{m,n-1}                                      \nn\\
&& +   \frac{g \sqrt{n-2} \sqrt{n-1} \sqrt{n} \left(g^2 \left(3219 n^2-6438 n+4384\right)-312 g (n-1)+32\right)  }{128 \sqrt{2}}          \  \delta_{m,n-3}                                      \nn\\
&& + \frac{g^2 \sqrt{n-4} \sqrt{n-3} \sqrt{n-2} \sqrt{n-1} \sqrt{n} (73 g (n-2)-4)  }{64 \sqrt{2}}                                              \  \delta_{m,n-5}     \nn\\
&& +   \frac{g^3 \sqrt{n-6} \sqrt{n-5} \sqrt{n-4} \sqrt{n-3} \sqrt{n-2} \sqrt{n-1} \sqrt{n}  }{64 \sqrt{2}}                                            \  \delta_{m,n-7}     ~~\label{Nxmn}                        
\ee
In the case of  $n>1$   we can write a general relation $
   {x_{mn}}  \approx    \sqrt n\ \sum_i\( \alpha_{i0} + \alpha_{i1}\cdot  (g   n) + \alpha_{i2}\cdot (g n)^2\)\ \delta_{n,m+i}$ 
Comparing this relation to  $  {E_{m }} $  in (\ref{crucial})   we see that the property of   “enhancement"  also shows in the  matrix elements  $   {x_{mn}} $.
\section{Third-order Perturbative OTOC of Anharmonic Oscillator : Numerical   $C_T(t)$ and Log$[C_T(t)]$}
To proceed, we substitute the analytic form  of $ {x_{mn}}$ in (\ref{Nxmn}) to calculate the  function $b_{nm}$ in  (\ref{bnm}).  Then we use the formula  (\ref{cn}) to calculate the associated  microcanonical OTOC $c_{n}(t)$.   Finally, we apply the formula   (\ref{TC}) to numerically evaluate the thermal  OTOCs, $C_T(t)$ and Log$[C_T(t)]$.  
\subsection{Thermal OTOC   $C_T(t)$ and  Log$[C_T(t)]$ : Scrambling and Saturation}
We plot  Figure 2 to show the general properties of $C_T(t)$ in the system with g=0.005 at T=60.  
\\
\\
\scalebox{0.6}{\hspace{2cm}\includegraphics{fig2}}
\\
{Figure 2:  Third-order   OTOC   $C_T(t)$ and Log$[C_T(t)]$ in the system with g=0.005 at T=60  as a function of time.}
\\

Figure 2  describes the time dependence of OTOC in the case with quartic interaction strength  g=0.005  at a temperature   T=60. The OTOC initially increase   and then   begins to oscillate. From the curve, we observe that the oscillation behavior  changes after  $t\approx 2$, where we mark with a red line. This difference leads us to conclude that   within the interval $0.6\le t\le 2$ the OTOC is scrambling. After scrambling  it  exhibits a fluctuation around a saturation point at later times. 

Note that the oscillation in the saturation region shown in the figure is the result of performing fine mode summation and finite order perturbation.  It is expected that, with more higher-order corrections and the inclusion of additional higher-mode summations, the oscillations will be smaller, as can be seen when comparing Figure 1, the results of second order, with Figure 2, the results of third order.
\subsubsection{Saturation Property}
 It is expected that, after the Ehrenfest time,   the thermal OTOC $C_T(t)$ will asymptotically in time to become $C_T(\infty)\to 2\<x^2\>_T \<p^2\>_T$ \cite{Maldacena16}, which is associated with the quantum chaotic behavior in systems that exhibit chaos. To calculate this quantity we can use the analytical method in section III to derive  the following third-order relations
\be
 {\<x^2\>}&=&(n+\frac{1}{2} )-\frac{3}{2} g \left(2 n^2+2 n+1\right)+\frac{5}{8} g^2 \left(34 n^3+51 n^2+59 n+21\right)\nn\\
&&-\frac{3}{2}   g^3 \left(125 n^4+250 n^3+472 n^2+347 n+111\right)   \label{A2} 
\\
  { \<p^2\>}&=&(n+\frac{1}{2})+\frac{3}{2} g \left(2 n^2+2 n+1\right)  -\frac{3}{8} g^2 \left(34 n^3+51 n^2+59 n+21\right)\nn\\
&&+ \frac{3}{4} g^3 \left(125 n^4+250 n^3+472 n^2+347 n+111\right) \label{A3}       
\ee
which reduces to $\<x^2\>=\<p^2\>=n+\frac{1}{2}$ if g=0 as it shall be. Then, using the above relations to do numerical mode summation to find the associated thermal average $2\<x^2\>_T \<p^2\>_T$. Comparing it to the  asymptotic value  of the thermal OTOC, Log[$C_T(\infty)]$,  calculated in Figure 2,  we can see that they  are quite similar to each other,  as was already found in the second-order approximation \cite{Huang2311}.  
\subsubsection{Mode Summation Constraint} \label{mode}
 It is easy to see that eq.(\ref{A2}) will become negative if n is too large\footnote{This   relates to the Enhancement Property described in Sec.\ref{Enhancement}.}. In this case the perturbation is invalid because $\<x^2\>$  should be definitively positive. Therefore, mode summation   should be subject to the constraint : $n<n_{cut}$, in which   $ n_{cut}$   depend on the  coupling constant g as was discussed  in the second-order approximation \cite{Huang2311}.  The figure 2 describes g=0.005 system and  is plotted with $n_{cut}=60$ therefore.

\subsection{Lyapunove Exponent in the Early Stage}
We plot Figure 3 to show more detailed properties of $C_T(t)$  in the early stage.

\scalebox{0.6}{\hspace{1cm}\includegraphics{fig3}}
\\
{Figure 3:  Third-order   OTOC   $C_T(t)$ and Log$[C_T(t)]$ in the system with g=0.005 and T=60 as a function of time in the early stage. The property of  exponential growth is shown in the initial stage.}
\\
 
The left-hand side of Figure 3 shows that the OTOC will  increase initially and then turns to  grow exponentially in the long interval $0.2< t <0.6$.   The function Log$[C_T(t)]$, plotted on the right-hand side of Figure 3, is nearly a linear function of time t during the long interval before the scrambling. In this way, the associated Lyapunov exponent can be determined from the slope of the red straight line in the Figure.    The properties shown in Figures 2 and 3 are also observed in the system with various values of g  and T.
\subsubsection{Statistical Analysis of  Exponential Growth}
 To see how the  evidence for the claimed exponential growth behavior let us present the statistical analysis for the function Log$[C_T(t)]$ in the systems with  g=0.005 at T=60.  Below, we present several fitting functions and their associated standard errors :

First, we fit $C_T(t)$  with   various functions  in the interval straight line  $0.1<t<0.4$ and calculate  the associated standard error   :
\be
&&  C_T(t) =a+b\ t , ~~~~~~~~~~~~~~~ \text{standard error} = 0.0597            \\
&&  C_T(t) =a+b\ t^2 , ~~~~~~~~~~~~~~ \text{standard error} =  0.0309               \\
&&   C_T(t) =a+b\ t^4   , ~~~~~~~~~~~~~~ \text{standard error} =  0.16594              \\
&&     C_T(t) =a\ e^{b\ t}  , ~~~~~~~~~~~~~~~~~ \text{standard error} =  0.01966    
\ee
Above results show that the exponential growth is a better fitting  function then others.

Next,  we  fit Log$[C_T(t)]$ over several time intervals using an exponential function and calculate the associated standard error. :
\be
&&   0.2<t<0.4,~~~ ~~ \text{Log}[C_T(t)] =5.4725   + 5.9644  \ t  , ~~~~~~~ ~~~~~ \text{standard error} =  0.00528 \\  
&&   0.2<t<0.5,~~~ ~~ \text{Log}[C_T(t)] =5.7547   + 4.9642   \ t  , ~~~~~~~ ~~~~~ \text{standard error} = 0.00939  \\ 
&&   0.2<t<0.6,~~~ ~~ \text{Log}[C_T(t)] =5.9783   + 4.2531  \ t  , ~~~~~~~ ~~~~~ \text{standard error} =  0.01278   
\ee
Tth exponential growth  within the interval  $0.2\le t\le 0.4$ has standard error $ \approx 0.00528$, which is the smallest of all.  However, in figure 3  we adopt the interval  $0.2\le t\le 0.6$  to describe the exponential growth in the system with g=0.005.  The reason is that, as discussed in the next subsection, the oscillating exponential can become pure exponential growth when higher-order terms are included. Thus, the oscillating curve in the interval $0.4\le t\le 0.6$ will transition into exponential growth if higher order calculations are included in the system with g=0.005. In this way, the scrambling is within the interval  $0.6\le t\le 2$, and the system is in saturation phase after  $t>2$ as shown in the figure 2. 
\subsubsection{General Property at Initial Stage}
{\bf  Strength  of interaction : } 
 The property of exponential growth curves fitting long time windows in third-order perturbation analysis could be observed in many systems with other values of g, for example g=0.003, g=0.03, but not all.  For example, the  OTOC at small g, such as g=0.001 which is described in figure 3 and figure 4, will exhibit oscillating exponential in its initial phase and  lacks the  visible characteristic straight line (on a log-linear plot) expected during an exponential growth phase\footnote{The property of oscillating exponential   is confirmed by the statistical analysis, in which $\text{Log}[C_T(t)]= 6.48851   + 0.22447   \ t$ with standard error 0.01042 in the interval $4<t<10$.}.  The reason may be traced to the fact that for small coupling system the initial stage property is dominated  by the quadratic coupling and thus the OTOC is oscillating like a SHO system  while the enhancement property could lead to the saturation in the final stage even the coupling is small. 
\\
\scalebox{0.6}{\hspace{2cm}\includegraphics{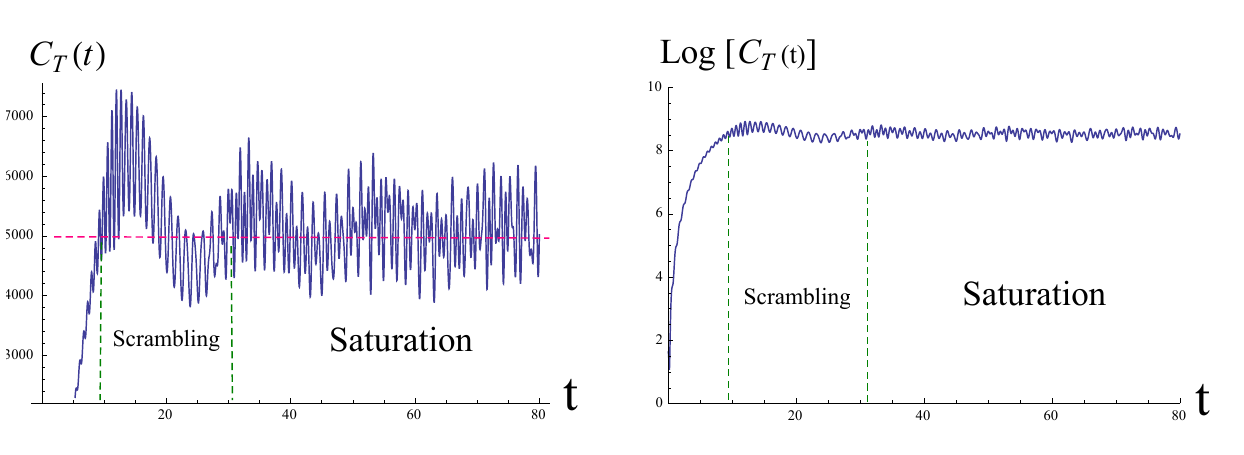}}
\\
{Figure 4:  Third-order   OTOC   $C_T(t)$ and Log$[C_T(t)]$ in the system with g=0.001 at T=60  as a function of time.}

\scalebox{0.6}{\hspace{1cm}\includegraphics{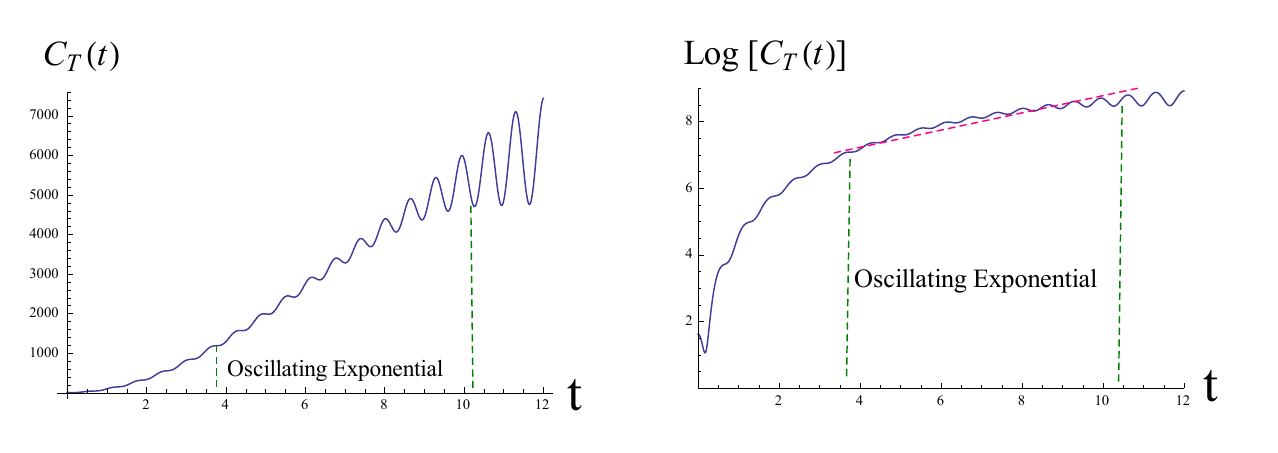}}
\\
{Figure 5:  Third-order   OTOC   $C_T(t)$ and Log$[C_T(t)]$ in the system with g=0.001 and T=60 as a function of time in the early stage. The property of oscillating exponential is shown in the initial stage.}
\\
\\
{\bf  Order of perturbation : } 
The oscillating exponential shown in Figures 4 and 5 can be observed in the second-order OTOC of the system with g=0.005. Notice that the third-order OTOC for the same system exhibits exponential growth, as illustrated in Figures 2 and 3. It is difficult to determine a precise critical value of  g or a critical order at which exponential growth emerges. However, we can be confident that higher orders of perturbation or a stronger quartic potential will lead the system to exhibit exponential growth behavior.
\\

In conclusion, in systems with sufficiently large quartic coupling and/or a high perturbation order, the OTOC can exhibit exponential growth before the scrambling phase.
\subsubsection{Concave Curve and Quartic Power Law in the Initial Stage}
We make one more comment  to discuss the OTOC properties in the early stage.   Our analytic formula   give the  OTOC at initial time  gives
\be
 \text{Log$[C_T(t)]$}=\alpha-\beta\cdot t^2  +\gamma \cdot t^4 +...  
\ee
where $\alpha~\beta$  and $\gamma$ are positive values which depend on the coupling $g$ and temperature $T$. The relative sign in the first two terms explains the behavior of concave curve in the right-hand figure nearly t=0.  The positive term linear in $t^4$ means that the OTOC will then raise in the quartic power law.  Note that a quartic power law function in Log$[C_T(t)]$  was also observed in previous literature \cite{Romatschke} through numerical calculation in wavefunction approach. Notice that  \cite{Romatschke}   considered pure quartic potential and find the quadratic power law function while in this paper we consider the simple harmonic oscillator coupled to quartic potential and find the  phase of   exponential growth curve fits a long time window.
 \section{Conclusions}
In this paper, we use the method established by Hashimoto recently in  \cite{Hashimoto17,Hashimoto20a,Hashimoto20b} to study the OTOC in the  quantum harmonic oscillator with an additional  quartic interaction.  We  calculate  the OTOC  by second quantization method in a three-order  perturbative approximation, which could provide us with several useful analytical relations. 

With the help of the analytical relations,  we apply the formula in (\ref{TC}) to numerically calculate the thermal  OTOC, $C_T(t)$.    We plot some diagrams to show that the OTOC in systems with sufficiently strong quartic interactions, the exponential growth curve fitting over a long time window emerges clearly in third-order perturbation. Then,  after scrambling,  the system  becomes a fluctuation around a saturation point in the final times.  This explicitly demonstrates the quantum chaos property in the harmonic oscillator with the extra quartic interaction.
 \\

Finally, we mention three  further studies relevant to our  research. 

$\bullet$ Firstly, the out-of-time-order correlator (OTOC) of a harmonic oscillator with quartic interaction was first studied using perturbation approximation from a second quantization approach in our series of papers. It would be interesting to study the system using the wavefunction approach.

$\bullet$ Secondly, the OTOC of non-linearly coupled oscillators was examined in an intriguing paper \cite{Hashimoto20a}. The properties of early-time exponential growth of the OTOC and the saturation property at late times, which are observed in systems exhibiting quantum chaos, were identified. However, a first-order perturbation analysis in an unpublished note did not observe these features  \cite{Huang2306},  suggesting that higher-order perturbations may be necessary to uncover them.

$\bullet$ Thirdly, the problem of many-body chaos at weak coupling was investigated several years ago by Stanford \cite{Stanford}, focusing on the system of matrix   $\Phi^4$ theory. In a recent paper, Kolganov and Trunin provided a detailed study of the classical and quantum butterfly effect in a related theory \cite{Kolganov}. Given that the interacting scalar field theory can be transformed into a system of coupled oscillators (see, for example, \cite{Huang21}), it is interesting to study the problems along the prescription of this paper.
\appendix
\section{Function  Form  of $|n\>^{(3)}$}
\be
&&|n\>^{(3)}~~~\nn\\
& & = \frac{1}{24576}\sqrt{n-11} \sqrt{n-10} \sqrt{n-9} \sqrt{n-8} \sqrt{n-7} \sqrt{n-6} \sqrt{n-5} \sqrt{n-4} \sqrt{n-3} \sqrt{n-2} \sqrt{n-1} \sqrt{n}                                \ |n-12\>^{(0)}\nn\\
& &   +  \frac{1}{6144}     \sqrt{n-9} \sqrt{n-8} \sqrt{n-7} \sqrt{n-6} \sqrt{n-5} \sqrt{n-4} \sqrt{n-3} \sqrt{n-2} \sqrt{n-1} \sqrt{n} (6 n-19)                                              \ |n-10\>^{(0)}\nn\\
& &   +   \frac{1}{ 768 }   \sqrt{n-7} \sqrt{n-6} \sqrt{n-5} \sqrt{n-4} \sqrt{n-3} (n-2)^{3/2} \sqrt{n-1} \sqrt{n} (6 n-31)                                                \ |n-8\>^{(0)}\nn\\
& &   + \frac{1}{ 6144 }     \sqrt{n-5} \sqrt{n-4} \sqrt{n-3} \sqrt{n-2} \sqrt{n-1} \sqrt{n} (n (n (122 n-2283)+6217)-5466)                                                \ |n-6\>^{(0)}\nn\\  
& &  -    \frac{1}{24576 }\sqrt{n-3} \sqrt{n-2} \sqrt{n-1} \sqrt{n} (n (n (n (387 n+23278)-112959)+166670)-98496)                                                   \ |n-4\>^{(0)}\nn\\ 
& &   +  \frac{1}{3072  }  \sqrt{n-1} \sqrt{n} (n (n (n (n (1175-198 n)+35372)-40127)+56650)-14412)                                               \ |n-2\>^{(0)}               \nn\\   
& &   +  \frac{3}{256  }  (2 n+1) (n (n+1) (89 n (n+1)+970)+744)         \ |n \>^{(0)}\nn\\  
& &   + \frac{1}{  3072} \sqrt{n+1} \sqrt{n+2} (n (n (n (n (198 n+2165)-28692)-137213)-237330)-145188)                                                    \ |n+2\>^{(0)}\nn\\  
& &   +\frac{1}{24576  }  \sqrt{n+1} \sqrt{n+2} \sqrt{n+3} \sqrt{n+4} (n (n (n (387 n-21730)-180471)-460874)-401016)                                                 \ |n+4\>^{(0)}\nn\\
& &   -\frac{1}{ 6144 }   \sqrt{n+1} \sqrt{n+2} \sqrt{n+3} \sqrt{n+4} \sqrt{n+5} \sqrt{n+6} (n (n (122 n+2649)+11149)+14088)                                                   \ |n+6\>^{(0)}\nn\\
& &   - \frac{1}{768  }   \sqrt{n+1} \sqrt{n+2} (n+3)^{3/2} \sqrt{n+4} \sqrt{n+5} \sqrt{n+6} \sqrt{n+7} \sqrt{n+8} (6 n+37)                                                   \ |n+8\>^{(0)}\nn\\
& &   -\frac{1}{6144  }   \sqrt{n+1} \sqrt{n+2} \sqrt{n+3} \sqrt{n+4} \sqrt{n+5} \sqrt{n+6} \sqrt{n+7} \sqrt{n+8} \sqrt{n+9} \sqrt{n+10} (6 n+25)                                                    \ |n+10\>^{(0)}\nn\\
& &   - \frac{1}{24576  }    \sqrt{n+1} \sqrt{n+2} \sqrt{n+3} \sqrt{n+4} \sqrt{n+5} \sqrt{n+6} \sqrt{n+7} \sqrt{n+8} \sqrt{n+9} \sqrt{n+10} \sqrt{n+11} \sqrt{n+12}       \ |n+12\>^{(0)}  \nn\\   \label{3}
\ee
\\
\\
\begin{center} 
{\bf  \large References}
\end{center}
 
 \end{document}